\documentclass[11pt]{article}
\usepackage{amssymb}
\usepackage{epsfig}
\newlength{\bredde}
\def\slash#1{\settowidth{\bredde}{$#1$}\ifmmode\,\raisebox{.15ex}{/}
\hspace*{-\bredde} #1\else$\,\raisebox{.15ex}{/}\hspace*{-\bredde} #1$\fi}
\newcommand{\be}{\begin{equation}}
\newcommand{\ee}{\end{equation}}
\newcommand{\bea}{\begin{eqnarray}}
\newcommand{\eea}{\end{eqnarray}}
\newcommand{\nn}{\nonumber}

\newcommand{\la}{\lambda}

\newcommand{\sect}[1]{\setcounter{equation}{0}\section{#1}}

\begin{document}
\topmargin -1.4cm
\oddsidemargin -0.8cm
\evensidemargin -0.8cm
\title{\Large{{\bf 
Distributions of Dirac Operator Eigenvalues}}}

\vspace{1.5cm}
\author{~\\{\sc G.~Akemann}$^1$ and {\sc P.~H.~Damgaard}$^{2}$
\\~\\
$^1$Service de Physique Th\'eorique, CEA/DSM/SPh$\!$T Saclay\\
Unit\'e associ\'ee CNRS/SPM/URA 2306\\
F-91191 Gif-sur-Yvette Cedex, France\\~\\
$^2$The Niels Bohr Institute, Blegdamsvej 17\\ 
DK-2100 Copenhagen {\O}, Denmark}

\date{}
\maketitle
\vfill
\begin{abstract}
The distribution of individual Dirac eigenvalues is derived by relating
them to the density and higher eigenvalue correlation functions.
The relations are general and hold for any gauge theory coupled to fermions
under certain conditions which are stated. As a special case, 
we give examples of the lowest-lying eigenvalue distributions 
for QCD-like gauge theories
without making use of earlier results based on
the relation to Random Matrix Theory.
\end{abstract}
\vfill

\begin{flushleft}
SPhT-T03/177
\end{flushleft}
\thispagestyle{empty}
\newpage

\renewcommand{\thefootnote}{\arabic{footnote}}
\setcounter{footnote}{0}

\sect{Introduction}\label{intro}

Consider the partition function of Yang-Mills theory coupled vector-like
to fermionic matter in a given representation of the gauge group.
Integrating out the $N_f$ 
fermions in the Euclidean partition function one obtains
the formal expression
\be
{\cal Z} ~=~ \int\![dA]~\prod_{f=1}^{N_f}\det[i\slash{D}(A) - m_f]\ 
\exp\left[-S[A]\right] ~,
\label{ZQCD}
\ee
where $S[A]$ is the pure gauge theory action. This involves the determinant
of the Dirac operator $\slash{D}$ (plus the mass terms). One way to make
the evaluation of this determinant meaningful is to regularize
the theory in such a way that there is a finite number of Dirac operator
eigenvalues $\lambda_i$. 

A regularization of the theory in such a way already exists:
lattice gauge theory with a finite lattice spacing $a$ and a finite
four-volume $V$. Because of chiral symmetry and the index theorem the
spectrum of the Dirac operator eigenvalues is constrained: the spectral
density $\rho(\lambda)$ will be symmetric around the origin, $\rho(\lambda)
= \rho(-\lambda)$, and there will, for any given gauge field configuration, 
be a number $\nu$ of exact zero modes. This number of zero modes
is dictated by the index theorem. 

What will be the distribution of
the remaining eigenvalues? The spectral density gives the overall
distribution of the unordered set of eigenvalues, but one would like
to ask more detailed questions.  Here the lattice gauge
theory approach offers a fruitful angle to the problem. It is thus
instructive to view the path integral over the gauge field $A_{\mu}(x)$
as the limit of a discrete sum over gauge field configurations. For
each gauge field configuration one will have a set of non-zero
Dirac operator eigenvalues $\lambda_i$, where, because of the spectral
symmetry, only the $N$ positive ones need to be considered here. These
eigenvalues can be ordered according to their magnitudes,
$\lambda_1 < \lambda_2 < \ldots < \lambda_N$. Performing this ordering
configuration by configuration, one can ask for the individual
distribution of the $k$th Dirac operator eigenvalue. It is far
from obvious how one would go about computing such individual
distributions of ordered Dirac operator eigenvalues directly from first
principles and the quantized gauge field theory. In fact, it is surprising how
{\em little} apparently can be said from first principles. The only
rigorous statement we have been able to find in the literature
is the inequality \cite{VW}
\be
\lambda_1 \leq \frac{C}{L} ~,
\ee
where $\lambda_1$ is the smallest non-zero eigenvalue, $C$ is a
constant, and $L$ is
the linear extent of the four-volume. All this says is that the
smallest eigenvalue $\lambda_1$ approaches zero in the limit
$L \to \infty$ at a rate which is not slower than that of 
free fields. This
should be contrasted with what one obtains if one in addition assumes
spontaneous chiral symmetry breaking. 
From the Banks-Casher
relation \cite{BC}
\be
\rho(0) = L^4\Sigma/\pi ~,
\ee
where $\Sigma$ is the chiral condensate, it follows 
that a non-vanishing $\Sigma$
requires the Dirac operator eigenvalues to accumulate at substantially
faster rate, yielding (with some other constant $\tilde{C}$) \cite{LS}
\be
\lambda_1 \leq \frac{\tilde{C}}{L^4} ~.\label{better}
\ee
This has been succinctly rephrased by Shuryak and Verbaarschot
\cite{SV}: Spontaneous chiral symmetry breaking and the required
accumulation of eigenvalues near the origin according to
(\ref{better}) suggests that a more appropriate object to study near 
$\lambda \sim 0$ is the so-called microscopic spectral density \cite{SV},
\be
\rho_s(\zeta) ~\equiv~ \frac{1}{\Sigma V}
\rho\left(\lambda=\frac{\zeta}{\Sigma V}\right) ~.
\label{rhomicro}
\ee 
Remarkably, it has turned out that this microscopic spectral
density is computable analytically using the relation to the appropriate
low-energy effective field theory, a chiral Lagrangian \cite{OTV}. 
What is perhaps more astonishing, the same results can be obtained
from universality classes of large-$N$ (chiral) Random Matrix Theories
\cite{SV,VZ,V,ADMN}. To be precise, using the relation to
effective field theory it has been shown how to derive the
microscopic spectral density $\rho_s(\zeta)$ \cite{OTV} and spectral two-point
function \cite{TV1} for the symmetry breaking class of QCD. 
For the two remaining symmetry breaking classes the corresponding 
effective field theories have been compared perturbatively to the 
microscopic spectral density from chiral Random Matrix Theory  \cite{TV}.
For the symmetry breaking class of QCD
the procedure for deriving the higher $k$-point
spectral correlation functions using effective field theory is completely
clear, and it is only computational tedium that has so far prevented the
calculation of higher $k$-point functions. 
Very compact expressions
for these spectral $k$-point functions are known if one accepts
the conjecture of the relation to Random Matrix
Theory, and the relevant generating functional which eventually should
be derived from field theory is also known explicitly \cite{SpV,AF}. 
But in this paper we will not rely on chiral Random Matrix
Theory at all, and therefore we shall not make use of these explicit
results, even though they are almost certainly in agreement in what one will
obtain from effective field theory. 

The purpose of this paper is to point out that even the distribution
of individual Dirac operator eigenvalues can be computed directly
from effective field theory. This may seem surprising, because
the concept of an ordered set of Dirac operator eigenvalues is
not obviously related to quantum field theory observables. But
following the program described in refs. \cite{OTV}, it is clear
how to proceed if one instead wishes to derive all spectral correlation
functions. Having all spectral correlation functions there
should be no more spectral information in the theory, and individual
eigenvalue distributions ought to follow. This is indeed the case.
In fact, the derivation will be completely general, and can be
phrased in terms of fundamental notions from probability theory and
certain assumptions regarding the regularized QCD partition function
which we will state explicitly below. In particular, the discussion
will not be restricted to the ``$\epsilon$-regime''
of QCD, although this clearly is an important special case. 
The counting of eigenvalues
from the origin can trivially be replaced by the counting
from any other arbitrary point in the spectrum.

\sect{Spectral Correlation Functions and Eigenvalue Distributions} 

To proceed, we need precise definitions of the spectral correlation
functions. We remind the reader that we consider a situation in
which we have a finite number of Dirac operator eigenvalues, and we
focus on the $N$ eigenvalues lying above zero. To even define spectral
correlation functions of these $N$ Dirac operator eigenvalues 
$\lambda_{i=1,\ldots,N}$ we must assume 
the existence of a joint probability distribution function
${\cal P}_N(\lambda_1,\ldots,\lambda_N)$. 
For simplicity we will here
assume that this function has support on the whole space of
${\mathbf R}_+^N$ (the restriction to a subset hereof is immaterial
for the discussion that follows, and is readily implemented).
We will also assume that ${\cal P}_N(\lambda_1,\ldots,\lambda_N)$
is symmetric under interchange of all arguments. 
We choose the normalization convention
\be
\int_0^{\infty}\!d\lambda_1\ldots d\lambda_N~ {\cal P}_N(\lambda_1,\ldots,
\lambda_N) ~=~ 1 ~.
\ee
The limit $N \to \infty$ is easily taken in all formulas that follow below.

The $k$-point density correlation function, which gives the 
probability density to find the $k$ variables at values $\lambda_1,
\ldots,\lambda_k$ is then
\be
\rho_k(\la_1, \ldots, \la_k) \ \equiv\ \frac{N!}{(N-k)!}
\int_0^\infty d\la_{k+1} \ldots d\la_N
\,{\cal P}_N(\la_1, \ldots, \la_N)\ .
\label{rhok}
\ee
In quantum field theory one is most used to considering just the one-point
function, the spectral density $\rho(\lambda)$. As follows, it is
obtained by integrating the joint probability density over $N-1$
variables, $viz.$
\be
\rho(\lambda) = N\int_0^{\infty}\! d\lambda_2\ldots d\lambda_N
{\cal P}_N(\lambda,\lambda_2,\ldots,\lambda_N) ~,
\ee
the factor of $N$ easily being understood as due to having arbitrarily
chosen to keep just the first entry fixed while integrating out the rest.
The existence of a joint probability density 
${\cal P}_N(\lambda_1,\ldots,\lambda_N)$ is implicit in all work that
considers the spectral density and higher spectral correlation
functions of the Dirac operator in QCD and related field theories. 
In particular, the $N$-point function is 
simply proportional to the 
joint probability distribution function itself,
$\rho_N(\lambda_1, \ldots, \lambda_N)=N!\,{\cal P}_N
(\lambda_1, \ldots, \lambda_N)$.

It is useful to also define a $k$-th ``gap probability''. 
For simplicity, let us restrict ourselves to a gap adjacent to the
origin; it is straightforward to pick any other fixed interval
on the support of the eigenvalues. Define
\be
E_k(s) \ \equiv\ \frac{N!}{(N-k)!}
\int_0^s d\la_1\ldots d\la_{k}\int_s^\infty d\la_{k+1}\ldots d\la_N 
\,{\cal P}_N(\la_1, \ldots, \la_N) \ , \ \ \mbox{for}\ k=0,1,\ldots,N\ \ .
\label{Ek}
\ee
For $k=0$ this gives the probability that all variables are larger
or equal to $s$, $\la_{i=1,\ldots,N}\geq s$, meaning that the 
interval $[0,s]$ is free of eigenvalues. For general $k$ the interval 
$[0,s]$ is occupied by $k$ eigenvalues and $[s,\infty)$ is 
occupied by $N-k$ eigenvalues. 
Our first step consists in proving that the $k$-th gap probability 
can be written as a sum of integrals over $(N-k+1)$ different $l$-point 
functions. We use the simple identity 
\be
(a-b)^j \ =\ \sum_{l=0}^j (-1)^l {j \choose l}
a^{j-l} b^l\ ,
\label{ab}
\ee
and choose $a=\int_0^\infty d\la$ and $b=\int_0^s d\la$ to replace all 
the $(N-k)$ integrals $\int_s^\infty d\la$ in eq. (\ref{Ek}) 
by $a-b$:
\bea
E_k(s) &=& \frac{N!}{(N-k)!}\int_0^s d\la_1\ldots d\la_{k} \nn\\
&&\times\sum_{l=0}^{N-k} (-1)^l {N-k \choose l}
\left(\int_0^\infty\right)^{N-k-l}
\left(\int_0^s\right)^l d\la_{k+1}\ldots d\la_N \,{\cal P}_N(\la_1, 
\ldots, \la_N)\nn\\
&=& \sum_{l=0}^{N-k} (-1)^l \frac{1}{l!} 
\int_0^s  d\la_1\ldots d\la_{k+l}\ \rho_{k+l}( \la_1,\ldots, 
\la_{k+l})\ .
\label{Ekrho}
\eea
Here we have used the invariance of ${\cal P}_N(\la_1, \ldots, \la_N)$ under 
permutations. The formula (\ref{Ekrho}) neatly expresses the gap probability
in terms of spectral correlation functions \cite{BF}. For the simplest
case of $k=0$ we get
\be
E_{0}(s)\ =\ 1\ -\ \int_0^s d\la_1\, \rho_1(\la_1)\ +\ \frac12 
\int_0^s d\la_1d\la_2\, 
\rho_2(\la_1,\la_2)\ - \frac{1}{6}\int_0^s d\la_1d\la_2d\la_3 
\rho_3(\la_1,\la_2,\la_3) +\ 
\ldots ~.
\label{E1example} 
\ee
We may also introduce a generating functional for all gap probabilities, 
defined as \cite{BF}
\be
E(s;\xi)\ \equiv\  1 + \sum_{l=1}^{N} (-\xi)^l \frac{1}{l!} 
\int_0^s  d\la_1\ldots d\la_{l}\ \rho_{l}(\la_1,\ldots, \la_{l}) \ .
\label{E}
\ee
It immediately follows that 
\be
E_k(s)\ =\ (-1)^{k} \left.\frac{\partial^{k}}{\partial\xi^{k}} \ 
E(s;\xi) \right|_{\xi=1} \ , \ \ \mbox{for}\ k=0,1,\ldots,N\ \ .
\label{Ekpartial}
\ee
The probability to find the $k$-th eigenvalue at value $\la_k=s$ is  
\be
p_k(s) \ \equiv\ k {N \choose k}
\int_0^s d\la_1\ldots d\la_{k-1}\int_s^\infty d\la_{k+1}\ldots d\la_N 
\,{\cal P}_N(\la_1, \ldots,\la_{k-1},\la_k=s,\la_{k+1},\ldots, \la_N)\ .
\label{pk}
\ee
If we order the eigenvalues, $\la_1 < \ldots < \la_N$, the quantity 
$p_k(s)$ gives the probability to find the $k$-th eigenvalue in this 
ordering at $\la_k=s$, 
as eq.(\ref{pk}) gives the probability to find $k-1$ eigenvalues 
in $[0,s]$, $\la_k=s$, and $N-k$ eigenvalues in the complement $[s,\infty)$. 
It is shown in Appendix \ref{A} that the distributions of eigenvalues $p_k(s)$
are indeed all properly normalized, $\int_0^\infty ds\ p_k(s)=1$ for all 
$k=1,\ldots,N$.

If we compare eq. (\ref{pk}) and eq. (\ref{Ek}) for $k-1$ 
we observe that they differ 
only by one integration. By differentiating we thus easily obtain
\be
 \frac{\partial}{\partial s} E_k(s)\ =\ k!\, \left( p_{k}(s) - 
p_{k+1}(s)\right)\ , 
\label{Ekpk}
\ee
defining $p_0(s)\equiv 0$. Here we have again made use of the permutation 
symmetry of the integrand, renaming the variable of the differentiated 
integrals by $\la_k$. Inserting eq. (\ref{Ekrho}) or eq. (\ref{Ekpartial})
we can thus recursively determine the distribution of the $k$-th 
individual eigenvalue from the knowledge of all density correlation
functions alone. This confirms the intuition mentioned in the
introduction that if all spectral correlations are known, no more
spectral information can be gleaned.

Let us give some examples. For the first eigenvalue we obtain 
\be
p_1(s)\ = \ -\frac{\partial}{\partial s} E_0(s)
\ =\ \rho_1(s) -\int_0^sd\la\, \rho_2(\la,s) \ + \ldots \ ,
\label{p1example}
\ee
by a single differentiation of eq. (\ref{E1example}),
whereas the second eigenvalue is given by 
\be
p_2(s)\ = \ \int_0^sd\la\, \rho_2(\la,s) \ +\ \ldots \ \ .
\label{p2example}
\ee
This explicitly confirms the expectation
that the first approximation to the smallest of these ordered eigenvalues
is given by the spectral density itself. The higher order correlation
functions systematically correct for the error in this initial
approximation. As we shall show below for the case of the microscopic
spectrum of the Dirac operator, the convergence can be remarkably
fast. Proceeding recursively, an infinite sequence of eigenvalue 
distributions follows from eq. (\ref{Ekpk}).

\sect{Distributions of Dirac Operator Eigenvalues in the 
$\epsilon$-Regime}

An obvious place to apply the above general formalism is near
the origin. As mentioned in the introduction, if one assumes that
chiral symmetry is broken spontaneously the spectral correlation
functions for the lowest-lying Dirac
operator eigenvalues can be computed analytically. The bridge is
an effective field theory which hinges on the precise Goldstone
boson manifold, and the pertinent field theory is taken to a corner
known as the $\epsilon$-regime \cite{GL} by a suitable tuning
of the four-volume $V$.
Let us mention that if we want to compare the regularized gauge 
theory with a finite number of eigenvalues $N$ to such an effective theory 
we must assume that this number $N$ is large.

The density correlation functions and thus 
also the distributions of eigenvalues depend on the number 
of flavors $N_f$, quark masses $m_f$ and gauge field topology $\nu$.
There have been several numerical simulations of individual
Dirac eigenvalue distributions both using staggered
fermions (see, $e.g.$, refs. \cite{Tilo,DHNR}) and overlap fermions
\cite{Urs,Wolfgang,Weisz}. So far it is has only been possible
to compare with the analytical predictions based on the conjectured
Random Matrix Theory results \cite{NDW,DN}, and an outstanding question
has been whether these individual eigenvalue distributions
also can be derived directly from the effective field theory. We will
here rely only on what to date has been derived from field theory, namely
the one- and two-point functions for the symmetry breaking class 
of QCD-like gauge theories (gauge theories with fermions
transforming according to complex representations of the gauge group)
\cite{OTV,TV1}.

\begin{figure}[-h]
\centerline{
\epsfig{figure=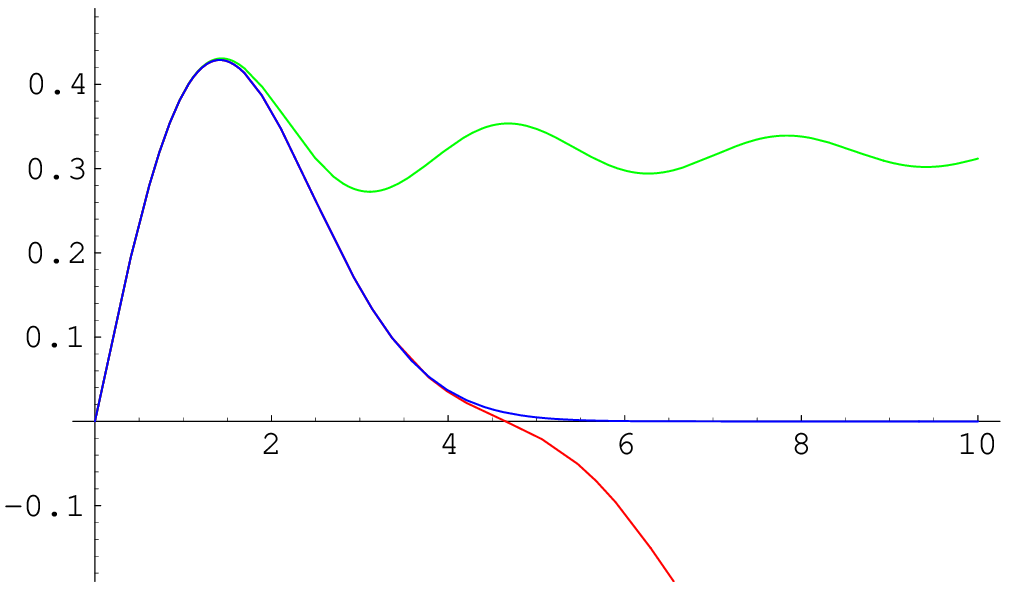,width=18pc}
\put(5,30){$x$}
\put(-240,120){$p_1^{(0)}(x)$}
\epsfig{figure=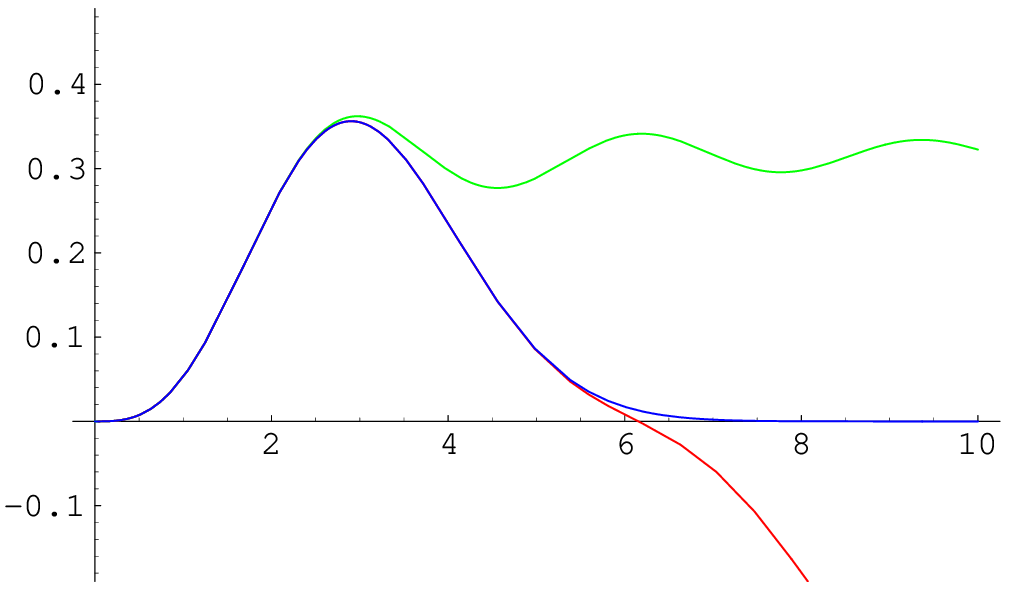,width=18pc}
\put(10,30){$x$}
\put(-240,120){$p_1^{(1)}(x)$}
}
\centerline{
\epsfig{figure=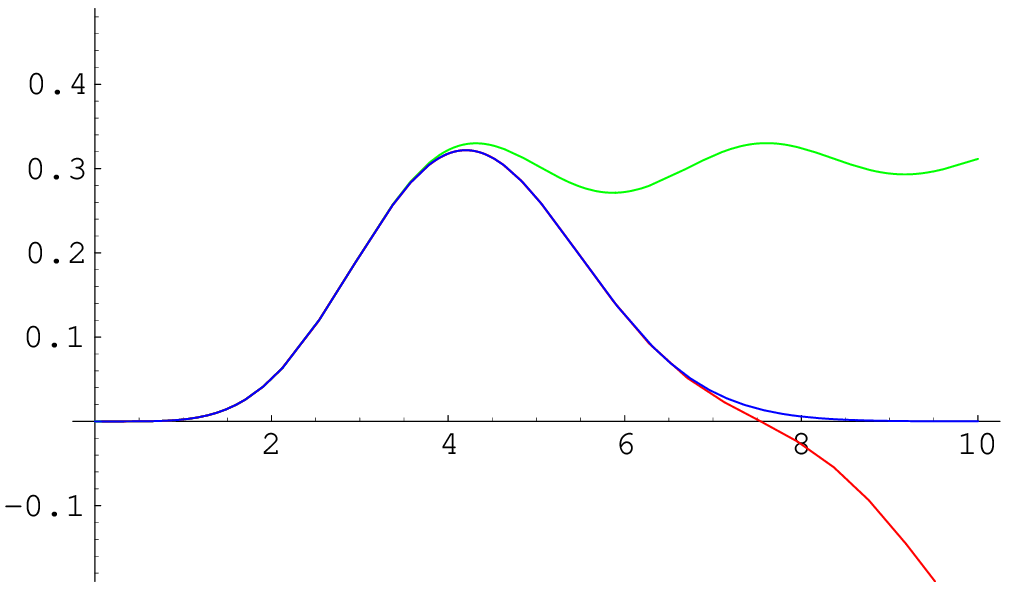,width=18pc}
\put(10,30){$x$}
\put(-240,120){$p_1^{(2)}(x)$}
}
\caption{The distribution of the first Dirac eigenvalue 
for $N_f=0$ and $|\nu|=0,1,2$.
The first term (green) and the sum of first and second term (red) 
in the expansion eq. (\ref{p1example})
are compared to the full result from Random Matrix Theory (blue) 
summing over all density correlations.}\label{p1}
\end{figure}
For simplicity of illustration we restrict ourselves to massless quarks.
The microscopic spectral density eq. (\ref{rhomicro}) 
for $N_f$ massless flavors in a sector of topological charge $\nu$
derived in \cite{OTV} reads
\be
\rho_s^{(N_f,\,\nu)}(\zeta)
\ =\ \frac{\zeta}{2} \left[ J^2_{N_f+|\nu|}(\zeta)
- J_{N_f+|\nu|+1}(\zeta)J_{N_f+|\nu|-1}(\zeta)\right]\ ,
\label{rho1}
\ee
in terms of the rescaled eigenvalues $\zeta=\la V\Sigma$. 
The two-point function has been derived so far only for zero flavors 
$N_f=0$ and arbitrary $\nu\neq0$ \
\be
\rho_s^{(N_f=0,\,\nu)}(\zeta,\eta)\ =\ 
\rho_s^{(0,\,\nu)}(\zeta)\rho_s^{(0,\,\nu)}(\eta)-
\frac{\zeta\eta}{(\zeta^2-\eta^2)^2} 
\left[ \zeta J_{|\nu|+1}(\zeta)J_{|\nu|}(\eta)
-\eta J_{|\nu|+1}(\eta)J_{|\nu|}(\zeta)\right]^2 ~.
\label{rho2}
\ee
We can now give the approximate 
distribution of the first Dirac Eigenvalue 
as it follows from the effective field theory to this order 
by inserting the above equations 
into eq. (\ref{p1example}). This is shown in Fig. \ref{p1} for three
different values of $\nu$. Clearly the approximation cannot
be trusted in the region where, to this order, $p_1^{(\nu)}(\zeta)$ goes
negative. 
However, the comparison to the full result as it is conjectured 
from Random Matrix Theory \cite{NDW} shows that for all practical purposes 
even this approximation that keeps only the two leading terms in the
expansion is sufficient when comparing with lattice data.

\begin{figure}[-h]
\centerline{
\epsfig{figure=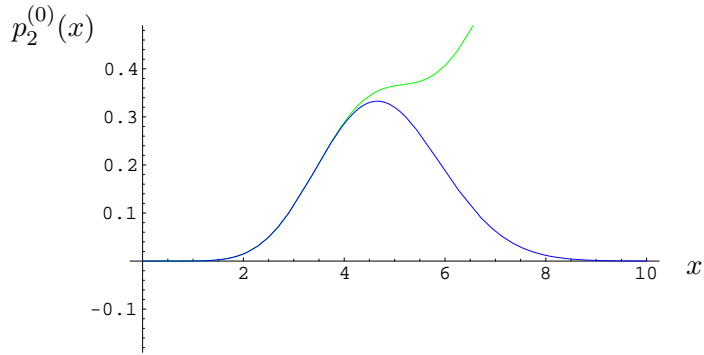,width=18pc}
\put(10,30){$x$}
\put(-245,120){$p_2^{(0)}(x)$}
}
\caption{The distribution of the second Dirac eigenvalue for 
$\nu=0$. The leading order term from eq. (\ref{p2example})
(green) is plotted versus the full result (blue).}
\label{p2}
\end{figure}
From eq. (\ref{p2example}) and the knowledge of the two-point function 
eq. (\ref{rho2}) we can also give the leading term in the expansion for 
the second Dirac eigenvalue, see Fig. \ref{p2}.
This 2nd eigenvalue distribution is described very accurately almost up to its 
maximum if we compare to Random Matrix Theory \cite{DN}. To get the
same degree of accuracy as for the first eigenvalue we would need
the three-point function, which has not yet been derived from the
chiral Lagrangian\footnote{We have checked this explicitly by using the 
three-point function conjectured from Random Matrix Theory.}. 
But we hope these examples suffice to illustrate how
individual lowest-lying eigenvalue distributions of the Dirac operator
can be obtained from the effective Lagrangian framework.

\sect{Conclusions}

We have shown that the distribution of individual Dirac operator 
eigenvalues can be computed solely from the knowledge of eigenvalue density 
correlations function. Since a well-established procedure exists
for deriving all $k$-point spectral correlation functions from
field theory,
this settles the question as to how one can also 
derive from field theory individual eigenvalue distributions of the Dirac 
operator. We have illustrated the method with a case where 
exact analytical results already have been derived from field theory,
namely the lowest-lying Dirac eigenvalues in QCD-like theories which
undergo spontaneous chiral symmetry breaking.
While it is expected that all eigenvalue correlations in this regime 
are identical to those obtained from Random Matrix Theory we have not 
made use of this assumption. We have shown that even just on the basis 
of the knowledge, to date, 
of the microscopic one- and two-point density from 
effective field theory the distribution of the first eigenvalue 
is indistinguishable from the conjectured analytical result from
Random Matrix Theory. The expansion can be pushed to any needed
degree of accuracy by computing the required higher
$k$-point correlation functions from field theory. This is
certainly the case for a physical number of fermions $N_f$.
The only exception may be the quenched limit of $N_f=0$. If this case is
inherently ill-defined at the level of the expansion in
the effective Lagrangian, not
even spectral correlation functions can be reliably computed by
the route through effective field theory. An entirely different approach
could then be called for. But our relations 
between individual eigenvalue distributions and spectral correlation
functions will remain valid.

\indent

\noindent
\underline{Acknowledgments}:
We wish to thank the CERN Theory Division, where part of this 
work was being done, for hospitality. The work of GA was supported by 
a Heisenberg fellowship of the Deutsche Forschungsgemeinschaft.

\begin{appendix}

\sect{Appendix A}\label{A}

Here we show that the distribution of the $k$-th eigenvalue 
eq. (\ref{pk}) is normalized to unity for all $k$:
\bea
\label{step1}
&& \int_0^\infty ds\, p_k(s) \ =\\ 
&=& k {N \choose k} \int_0^\infty ds
\int_0^s d\la_1\ldots d\la_{k-1}\int_s^\infty d\la_{k+1}\ldots d\la_N 
\,{\cal P}_N(\la_1, \ldots,\la_{k-1},s,\la_{k+1},\ldots, \la_N) \nn\\
&=& k {N \choose k} \sum_{l=0}^{k-1} 
(-1)^l {k-1 \choose l} 
\int_0^\infty ds
\left( \int_0^\infty \right)^{k-l-1}
\left( \int_s^\infty \right)^{N-k+l} \!\!
d\la_1\ldots d\la_{k-1}d\la_{k+1}\ldots d\la_N \nn\\
&&\ \ \ \ \ \ \ \ \ \ \ \ \ \ \ \ \ \ \ \ \ \ \ \ \ \ \ \ 
\ \ \ \ \ \ \ \ \ \ \ \ \ \ \ \ \ \ \ \ \ \ \ \ \ \ \ \ \ \ \ \ \ \times \ 
\,{\cal P}_N(\la_1, \ldots,\la_{k-1},s,\la_{k+1},\ldots, \la_N)\ . \nn
\eea
We have used again the identity eq. ({\ref{ab}}) choosing 
$a=\int_0^\infty d\la$ and $b=\int_s^\infty d\la$ 
in order to replace the integrals $\int_0^sd\la$. We next order 
the integration variables of the $(N-k+l)$ integrations over $[s,\infty)$, 
\bea
\label{step2}
&& \int_0^\infty ds\left( \int_s^\infty \right)^{N-k+l} 
d\la_{k-l+1}\ldots d\la_N \,
{\cal P}_N(\la_1, \ldots,\la_{k-1},s,\la_{k+1},\ldots, \la_N)
\ =\\
&=&(N-k+l)! \int_0^\infty ds \int_s^\infty d\la_{k-l+1}
\int_{\la_{k-l+1}}^\infty  d\la_{k-l+2}\ldots 
\int_{\la_{N-1}}^\infty d\la_N 
\nn\\
&&\ \ \ \ \ \ \ \ \ \ \ \ \ \ \ \ \ \ \ \ \ \ \ \ \ \ \ \ 
\ \ \ \ \ \ \ \ \ \ \ \ \ \ \ \ \ \ \ \ \ \ \ \ \ \ \ \ \ \ \ \ \ \times \ 
\,
{\cal P}_N(\la_1, \ldots,\la_{k-1},s,\la_{k+1},\ldots, \la_N) \nn\\
&=& \frac{1}{N-k+l+1} \left( \int_0^\infty \right)^{N-k+l+1}  
ds \,d\la_{k-l+1}\ldots d\la_N \,{\cal P}_N(\la_1, \ldots,\la_{k-1},s,
\la_{k+1},\ldots, \la_N) \ ,
\nn
\eea
and then undo the ordering with the additional integration over $s$. 
Inserting eq. (\ref{step2}) into eq. (\ref{step1}) all the integrations 
over the joint probability distribution will give unity and we arrive at
\be 
\int_0^\infty ds\ p_k(s) \ =\  
k {N \choose k} \sum_{l=0}^{k-1} 
(-1)^l {k-1 \choose l}  \frac{1}{N-k+l+1} \ =\ 1.
\ee

\end{appendix}

\end{document}